\documentclass[pra,aps,twocolumn,amsmath,amssymb,superscriptaddress]{revtex4}
\usepackage{bm,graphicx,amsmath}
\usepackage{bbm}

\usepackage{bm}
\usepackage{amssymb}
\usepackage{epsfig}
\usepackage{epstopdf}
\usepackage{amstext}
\usepackage{latexsym}
\usepackage{hyperref}
\usepackage{times}
\usepackage{color}

\begin{document}

\title{Entanglement control in hybrid optomechanical systems}

\author{B. Rogers}
\affiliation{Centre for Theoretical Atomic, Molecular and Optical Physics, Queen's University Belfast, Belfast BT7 1NN, United Kingdom}
\author{M. Paternostro}
\affiliation{Centre for Theoretical Atomic, Molecular and Optical Physics, Queen's University Belfast, Belfast BT7 1NN, United Kingdom}
\affiliation{Institut f\"ur Theoretische Physik, Albert-Einstein-Allee 11, Universit\" at Ulm, D-89069 Ulm}
\author{G. M. Palma}
\affiliation{NEST Istituto Nanoscienze-CNR and Dipartimento di Fisica, Universit\`a degli Studi di Palermo, via Archirafi 36, I-90123 Palermo, Italy}
\author{G. De Chiara}
\affiliation{Centre for Theoretical Atomic, Molecular and Optical Physics, Queen's University Belfast, Belfast BT7 1NN, United Kingdom}

\begin{abstract}
We demonstrate the control of entanglement in a hybrid optomechanical system comprising an optical cavity with a mechanical end-mirror  and an intracavity Bose-Einstein condensate (BEC). Pulsed laser light (tuned within realistic experimental conditions) is shown to induce an almost sixfold increase of the atom-mirror entanglement and to be responsible for interesting dynamics between such mesoscopic systems. In order to assess the advantages offered by the proposed control technique, we compare the time-dependent dynamics of the system under constant pumping with the evolution due to the modulated laser light.
\end{abstract}

\date{\today}

\maketitle

Over the last few years we have witnessed a constant-pace advance in the control of mesoscopic systems at the quantum level. Examples range from quantum interference experiments with large organic molecules~\cite{arndt} to the preparation and detection of multi-particle entanglement in superconducting devices~\cite{dicarlo}. All these efforts are contributing very significantly to the exploration of the quantum-to-classical boundary and the ascertainment of the existence of actual limitations in the enforcing of genuinely quantum mechanical behavior in mesoscopic systems. From a more applied viewpoint, these advances contribute to the quest for the manipulation of information encoded in open systems operating beyond the microscopic scale.

Cavity quantum optomechanics has preponderantly emerged, recently, as a very interesting arena for the study of quantum features at the meso-scale. Its aim is to control the quantum state of mechanical oscillators by their coupling to a light field~\cite{review}. Recent advances in this context include the realization of the {\it quantum-coherent coupling} of a mechanical oscillator with an optical cavity~\cite{verhagen} where the coupling rate exceeds both the optical and the mechanical decoherence rate, and the laser cooling of a nanomechanical oscillator to its ground state~\cite{chan}. When microwave radiation is used instead of optical fields, an analogous sideband cooling of the motion of a microscopic cantilever has been demonstrated, leading to a mechanical thermal occupation of  $\sim$1~\cite{teufel}.

Such new technologies open unprecedented possibilities in the design of hybrid quantum architectures whose elementary building blocks are physically implemented by systems of different nature. Achieving accurate control at their interfaces is quickly becoming of technologically and foundationally paramount importance. In this respect, the coherent coupling of a superconducting flux qubit to a spin ensemble has been recently reported~\cite{hybrid}, while architectures for the interaction between ultracold atoms and mechanical systems have been demonstrated~\cite{hunger}. Furthermore, recent experiments have reached the regime of strong coupling between a Bose-Einstein condensate (BEC) and an optical cavity~\cite{Esslinger2008,Stamper2008}, simulating optomechanical effects where the mechanical system is embodied by quantum phononic waves of the BEC. 

Here we consider a hybrid situation where we combine technology coming from the traditional optomechanical setting with the potential of the BEC-cavity experiments and proposing a new BEC-optomechanics apparatus. The original idea~ \cite{MauroPRL2010} is founded on the insertion of a BEC in an externally driven optical cavity whose end-mirror oscillates around an equilibrium position. For a small number of mechanical phonons, the composite system made of the cavity field, the BEC phononic mode, and the vibrating mirror is endowed with genuine multipartite entanglement that can in principle be revealed by all-optical measurements~\cite{DeChiara2011}. We propose the active driving control of such a hybrid system realized by time-modulating the intensity of the driving field. Inspired by recent works on the control of optomechanical devices~\cite{mari,schmidt}, we show that by using a monochromatic modulation, entanglement between two mesoscopic systems, the mirror and the BEC, can be created and controlled~\cite{farace}. Furthermore, by borrowing ideas from the theory of optimal control~\cite{doria} we show that, with respect to the unmodulated case, a sixfold improvement in the degree of generated entanglement  is in order. We interpret such performance in terms of the occurrence of a special resonance at which the building up of entanglement  is {\it favored}. Our results demonstrate the viability of the optimal control-empowered manipulation of open mesoscopic systems for the achievement of strong quantum effects, even in the hybrid context, of which the system that we study is a significant representative.

\section{The Physical Model}
We start describing the hybrid optomechanical  setup at hand~\cite{MauroPRL2010,DeChiara2011}, which consists of a Fabry-Perot cavity with a vibrating end-mirror. The cavity is pumped by a laser that couples to the mechanical mirror and an intracavity BEC.
The Hamiltonian of the system (in a frame rotating at the frequency $\omega_L$ of the pump field) reads
\begin{equation}
\label{eq:Hsum}
\hat{\cal H} = \hat{\cal H}_C + \hat{\cal H}_A + \hat{\cal H}_M + \hat{\cal H}_{AC} + \hat{\cal H}_{MC}.
\end{equation}
The Hamiltonian of the mirror is
\begin{equation}
\label{eq:HM}
\hat{\cal H}_M = \frac 12 m \omega_m^2 \hat{q}^2 +\frac{\hat{p}^2}{2m},
\end{equation}
where $m$ is the effective mechanical mass, $\omega_m$ is the free mechanical frequency and $\hat{q}$ ($\hat{p}$) is the position (momentum) operator of the mirror. The Hamiltonian of the driven cavity is 
\begin{equation}
\label{eq:HC}
\hat{\cal H}_C = \hbar (\omega_C{-}\omega_L)\hat{a}^\dagger\hat{a}{-}i \hbar \eta (\hat{a}-\hat{a}^\dagger),
\end{equation}
where $\omega_C$ is the cavity frequency, $\hat a$ is the cavity field's annihilation operator, and
$\eta = \sqrt{2\kappa{\cal R}/\hbar\omega_L}$ accounts for the laser pumping  [${\cal R}$ is the laser power and $\kappa$ is the cavity decay rate].
The BEC is taken to be weakly interacting, allowing the separation of the atomic field operator into a classical component (the condensate wave function) and a quantum one (the fluctuations), expressed in terms of Bogoliubov modes. The cavity is strongly coupled to the mode with wavelength $\lambda_c/2$ where $\lambda_c$ is the cavity-mode wavelength~\cite{Esslinger2008,MauroPRL2010}. We call $\omega_b$ the frequency of the Bogoliubov mode and $\hat{c}$ ($\hat{c}^\dagger$) the corresponding annihilation (creation) operator. As the condensate is at low temperature, any thermal fluctuation of the atoms is negligible. The free Hamiltonian of the Bogoliubov mode is given by
$\hat{\cal H}_A = \hbar \omega_b\hat{c}^\dagger\hat{c}$,
while the atom-cavity coupling is~\cite{Esslinger2008,MauroPRL2010}
\begin{equation}
\label{eq:HAC}
\hat{\cal H}_{AC}=\frac{\hbar g^2 N_0}{2\Delta_a}\hat{a}^\dagger\hat{a} + \hbar\sqrt{2}\zeta\hat{Q}\hat{a}^\dagger\hat{a},
\end{equation}
where $g$ is the atom-cavity coupling strength, $N_0$ is the condensate population, and $\Delta_a$ is the atom-cavity detuning.  In the second term, the atom-cavity coupling rate is denoted by $\zeta$. The complete derivation of its form can be found in Ref.~\cite{MauroPRL2010}. The position and momentum quadratures of the Bogoliubov mode are $\hat{Q}{=}(\hat{c}+\hat{c}^\dag)/\sqrt{2}$ and $\hat{P}{=}i(\hat{c}^\dag-\hat{c})/\sqrt{2}$, respectively.  Finally, the mirror-cavity interaction is given by
${\hat{\cal H}_{MC}{=}-\hbar \chi \hat{q}\hat{a}^\dagger\hat{a}}$
with $\chi=\omega_C/L$ the mirror-cavity coupling rate ($L$ is the length of the cavity).  As can be seen by comparing $\hat{\cal H}_{MC}$ with the second term of Eq.~(\ref{eq:HAC}), the BEC dynamics is analogous to a mechanical oscillator under the action of radiation pressure. As no direct coupling term ${\cal H}_{AM}$ is present in Eq.~(\ref{eq:Hsum}), the interaction between the atomic mode and the mirror is mediated by the cavity. The relevant degrees of freedom of the system are grouped in the vector 
$\hat{\cal \phi}^T = (\hat{x},\hat{y},\hat{q},\hat{p},\hat{Q},\hat{P})$,
where the cavity position and momentum-like quadrature operators are $\hat{x}=(\hat{a}+\hat{a}^{\dag})/\sqrt{2}$ and $\hat{y}=i(\hat{a}^{\dag}-\hat{a})/\sqrt{2}$, respectively.  Under intense laser pumping the operators can be linearised and expanded around their respective classical mean values $\phi_{s,i}$ such that $\hat{\phi_i} \rightarrow \phi_{s,i} + \delta\hat{\phi}_i$, where $\delta\hat{\cal \phi}^T = (\delta\hat{x},\delta\hat{y},\delta\hat{\tilde{q}},\delta\hat{\tilde{p}},\delta\hat{Q},\delta\hat{P})$ is the vector of zero-mean quantum fluctuations for each operator in $\hat{\cal \phi}$. Here, the mirror position and momentum operators have been rescaled to dimensionless quantities as $\hat{q}=\sqrt{\hbar/m\omega_m}\hat{\tilde{q}}$, and $\hat{p}=\sqrt{\hbar m\omega_m}\hat{\tilde{p}}$. In the hybrid optomechanical system considered here, two sources of noise should be taken into account. The first comes from photons leaking out of the cavity, while the second is due to the mechanical Brownian motion performed by the mirror, which is typically in contact with a thermal bath at temperature $T$. The open-system nature of the problem at hand allows for the establishment of a stationary state. In fact, the classical values $\phi_{s,i}$ can be calculated by solving the steady-state Langevin equations~\cite{Vitali2007}, which leads to the new equilibrium positions for the mechanical mirror and the harmonic oscillator embodied by the Bogoliubov mode
${q_s =\frac {\hbar \chi|\alpha_s|^2}{m\omega_m^2}}~~\textrm{and}, 
{Q_s = -\frac{\zeta|\alpha_s|^2}{\omega_b}}$.
%
Here we have introduced the mean intracavity field amplitude 
%
$|\alpha_s|^2 ={\eta^2}/({\Delta^2+\kappa^2})$
%
and the total cavity-pump detuning (modified by the radiation pressure mechanisms and the shift of the cavity frequency due to its off-resonant coupling with the atomic medium)
\begin{equation}
\label{eq:delta}
\Delta = \omega_C - \omega_L + \frac{g^2N_0}{2\Delta_a} - \chi' q_s + \zeta Q_s.
\end{equation}
As the last three terms in $\Delta$ are typically very small compared to the bare detuning $\omega_C - \omega_L$, we neglect any bistability effect and assume $\Delta$ to be an independent control parameter.

As for the fluctuations, their dynamics can be described by the following vector equation
\begin{equation}
\label{eq:timedphi}
\partial_t \delta{\hat{\bm \phi} }={{\cal K}}\delta\hat{\bm \phi}{+}\hat{\cal {\bm N}},
\end{equation}
where we have introduced the input-noise vector 
%
$\hat{\cal{\bm N}}^T =(\sqrt{2\kappa}\delta \hat{x}_{in},\sqrt{2\kappa}\delta \hat{y}_{in},0,{\hat{\xi}}/{\sqrt{\hbar m \omega_m}},0,0)$.
%
The drift matrix $\cal K$, given below, depends on the scaled coupling parameter $\chi{=}\chi'\sqrt{\hbar/m\omega_m}$ and the mirror dissipation rate $\gamma{=}\omega_m/Q$ [where $Q$ is the mechanical quality factor],
\begin{equation}
\cal{K} = \left(
\begin{array}{cccccc}
-\kappa & \Delta & 0 & 0 & 0 & 0 \\
-\Delta & -\kappa & \sqrt{2}\chi'\alpha_S & 0 & -\sqrt{2}\zeta\alpha_S & 0 \\
0 & 0 & 0 & \omega_m & 0 & 0 \\
\sqrt{2}\chi'\alpha_S & 0 & -\omega_m & -\gamma & 0 & 0 \\
0 & 0 & 0 & 0 & 0 & \Omega \\
-\sqrt{2}\zeta\alpha_S & 0 & 0 & 0 & -\Omega & 0 \\
\end{array}
\right).
\end{equation} 
The photon-leakage from the cavity is accounted for by the input-noise operators 
\begin{equation}
\delta\hat{x}_{in}=\frac{\delta\hat{a}^{\dag}_{in}+\delta\hat{a}_{in}}{\sqrt{2}}~~\textrm{and}~~
\delta\hat{y}_{in}=i\frac{\delta\hat{a}^{\dag}_{in}-\delta\hat{a}_{in}}{\sqrt{2}}
\end{equation}
with $\langle\delta\hat{a}_{in}\rangle=\langle\delta\hat{a}^\dag_{in}\rangle=0$ and $\langle\delta\hat{a}_{in}(t)\delta\hat{a}^{\dag}_{in}(t')\rangle=\delta(t-t')$. The Langevin force operator $\hat{\xi}$ in $\hat{\cal{\bm N}}$ models the effects of the mechanical Brownian motion. In the limit of a high mechanical quality factor, such noise can be faithfully considered as Markovian, as entailed by the asymptotic form of the auto-correlation function $\langle \xi(t) \xi(t') \rangle{\simeq}\hbar\gamma{m/\beta_B}\delta(t{-}t')$, where $\beta_B{=}\hbar/{2k_B T}$ and $k_B$ is the Boltzmann constant.

We consider a viable detection scheme to observe entanglement between the various bi-partitions.  Our proposal is linked to the method put forward in Ref.~\cite{DeChiara2011}, i.e. on the mapping of atom-field or mirror-field entanglement (and thus the atom-field one) into fully accessible all-optical quantum correlations by means of two extra fields that interact (locally) with the abovementioned subsystems. Both the operations are within reach of state-of-the-art experiments and, in fact, have been recently implemented~\cite{Groeblacher2009,Baumann2011} with high efficiency. Our revelation scheme will thus be affected by the limitations of such methods, which embody the forefront of weakly disruptive detection schemes in such mesoscopic scenarios.

\begin{figure}[t]
\includegraphics[width=\linewidth]{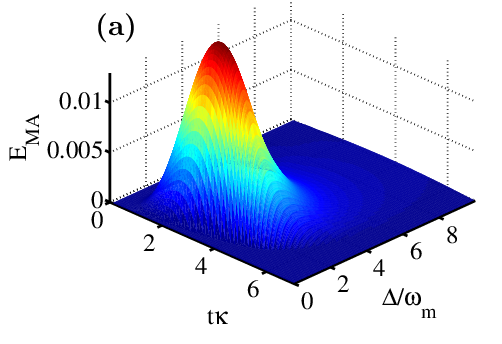}\\\includegraphics[width=0.98\linewidth]{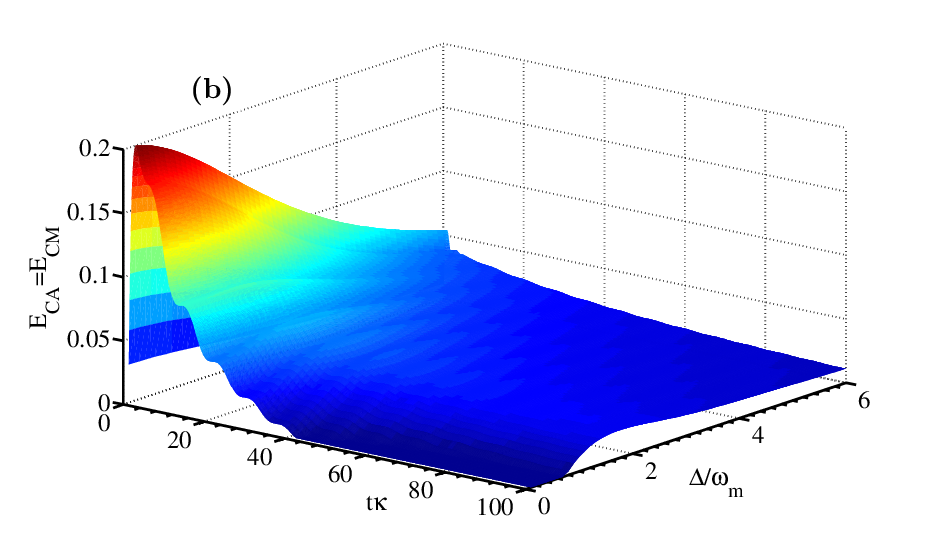}
\caption{(Color online) {\bf (a)} Entanglement $E_{MA}(t)$ between mirror and atoms against $\kappa t$  and  $\Delta/\omega_m$. {\bf (b)}  Same as panel {\bf (a)} for  $E_{CM}(t)=E_{CA}(t)$.
Parameters: $\omega_m/2\pi=3\times10^6~s^{-1}$; $T=10\mu K$; $Q=3\times10^4$; $m=50~\mathrm{ng}$; ${\cal R}=50~\mathrm{mW}$, cavity finesse $F=10^4$, and $\zeta=\chi$; cavity length is $L=1~\mathrm{mm}$ from which $\kappa=\pi c/2LF$ ($c$ is the speed of light).  All plotted units are dimensionless.}
\label{fig:deltaplots}
\end{figure}

\section{Entanglement dynamics}
In Ref.~\cite{DeChiara2011}, the stationary entanglement within the hybrid optomechanical system has been considered. Here we focus on the dynamical regime where the evolution of the entanglement is resolved in time.  We consider the fully symmetric regime encompassed by equal frequencies for the Bogoliubov and mechanical modes (i.e. $\omega_b=\omega_m$) and identical coupling strengths in the bipartite cavity-mirror and cavity-atom subsystems (that is, we take $\zeta=\chi$). We are particularly interested in the emergence of atom-mirror entanglement at short interaction times. The analysis conducted in Ref.~\cite{DeChiara2011} has shown it to be absent for $t\rightarrow\infty$. The linear nature of Eq.~\eqref{eq:timedphi} preserves the Gaussian nature of any initial state of the overall system. This allows us to fully characterise the entanglement evolution through the covariance matrix 
%
${\cal V}_{ij} = \frac{1}{2}\langle\{\delta\hat{\phi}_i,\delta\hat{\phi}_j\}\rangle{-}\langle\delta\hat{\phi}_i\rangle\langle\delta\hat{\phi}_j\rangle$.
%
Using this definition and Eq.~\eqref{eq:timedphi}, the dynamical equation regulating the evolution of the covariance matrix can be written as
\begin{equation}
\label{eq:Vdot}
\dot{{\cal V}} = {\cal KV}+{\cal VK}^T + {\cal D},
\end{equation}
where we have introduced the noise matrix ${\cal D} = \mathrm{diag}[\kappa, \kappa, 0, \gamma(2\bar{n}{+}1), 0, 0]$ with
$\bar{n}=[\mathrm{exp}(2\beta_B\omega_m){-}1]^{-1}$ as the thermal mean occupation number of the mechanical mode. In Eq.~(\ref{eq:Vdot}) we assume that the mean values of the mechanical and optical quadratures reach their stationary values much faster than the fluctuation dynamics (this is always verified in our calculations). Such an inhomogeneous first-order differential equation is solved assuming the initial conditions
${\cal V}(0) = \mathrm{diag}[{1},1,2\bar{n}{+}{1},2\bar{n}{+}{1},1,1]/2$, which describe the vacuum state of both the cavity field and the BEC mode and the thermal state 
of the mechanical system. Physicality of the covariance matrix has been thoroughly checked by considering the fulfillment of the Heisenberg-Robinson uncertainty principle and checking that the minimum symplectic eigenvalue $\nu=\min\mathrm{eig}(i\omega{\cal V})$ is such that $|\nu|\ge\frac{1}{2}$. Here used the $6 \times 6$ symplectic matrix $\omega=\oplus^3_{j=1}i\sigma_y$ with $\sigma_y$ the y-Pauli matrix. 
The entanglement measure that we use here to quantify entanglement between any two modes $\alpha$ and $\beta$ is the logarithmic negativity~\cite{vidal2002}, defined as $E_{\alpha\beta} = -\mathrm{log}|2\nu_{min}|$, where $\nu_{min}$ is the smallest symplectic eigenvalue of the matrix ${\cal V}_{\alpha\beta}^{T_\beta}=P{\cal V}_{\alpha\beta}P$, for $\alpha,\beta = C,A,M$ [the latter being the labels for the cavity, atomic and mirror modes, respectively].  The reduced covariance matrix ${\cal V}_{\alpha\beta}$ contains the entries of ${\cal V}$ associated with modes $\alpha$ and $\beta$ while, by inverting the momentum quadrature of $\beta$, matrix $P=\mathrm{diag}(1,1,1,-1)$ performs the partial transposition in phase space.
The atom-mirror entanglement $E_{MA}(t)$, whose time evolution is shown in Fig.~\ref{fig:deltaplots}{\bf (a)} for different values of the effective detuning $\Delta$, gradually develops and reaches its peak value as the cavity-atom and cavity-mirror entanglement drop to a quasi-stationary value.  As no direct atom-mirror interaction exists in this system, mediation through the cavity mode essentially results  in a delay: quantum correlations between the atoms (the mirror) and the cavity mode must build up before any atom-mirror correlation can appear. This is clearly shown in Fig.~\ref{fig:deltaplots}{\bf (b)}, where $E_{CM}(t)=E_{CA}(t)$ reach their maximum well before $E_{MA}(t)$ starts to grow.
The atom-mirror entanglement is non-zero only within a very short time window, signalling the fragility of quantum correlations resulting from only a second-order interaction between the BEC and the cavity end-mirror. These results go far beyond the limitations of the steady-state analysis conducted in~\cite{DeChiara2011} and prove the existence of a regime where all the various reductions obtained by tracing out one of the modes from the overall system are inseparable, a situation that, within the range of parameters considered in our investigation, is typical only of a time-resolved picture.

\subsection{Optimal control of the early-time entanglement}
We now consider the effects of time-modulating the external pump power ${\cal R}$, which is now considered a function of time. In turn, this implies that we now take $\eta\rightarrow\eta(t)$ and study the time behavior of the entanglement $E_{MA}$ set between the atoms and the mirror. We will show that a properly optimised $\eta(t)$ can increase the maximum value of $E_{MA}(t)$ for values of $t$ within the same time interval $\tau$ where atom-mirror entanglement has been shown to emerge in the unmodulated case.
We assume to vary $\eta(t)$ slowly in time, so that the classical mean values $\phi_s$ adiabatically follow the change in $\eta(t)$. This approximation is valid as long as the number of intra-cavity photons is large enough to retain the validity of the linearization procedure and the time-variation of $\eta(t)$ is slow compared to the time taken by the mean values to reach their stationary values. For all cases considered here we have verified the validity of such assumptions. The dynamics of the covariance matrix is thus still governed by Eq.~\eqref{eq:Vdot} with the replacement ${\cal K}\rightarrow{\cal K}(t)$. In the following, we use the value of $\Delta=2.7\omega_m$ which maximises the short time entanglement $E_{MA}$.

\begin{figure}[b]
\begin{center}
\includegraphics[width=0.99\linewidth]{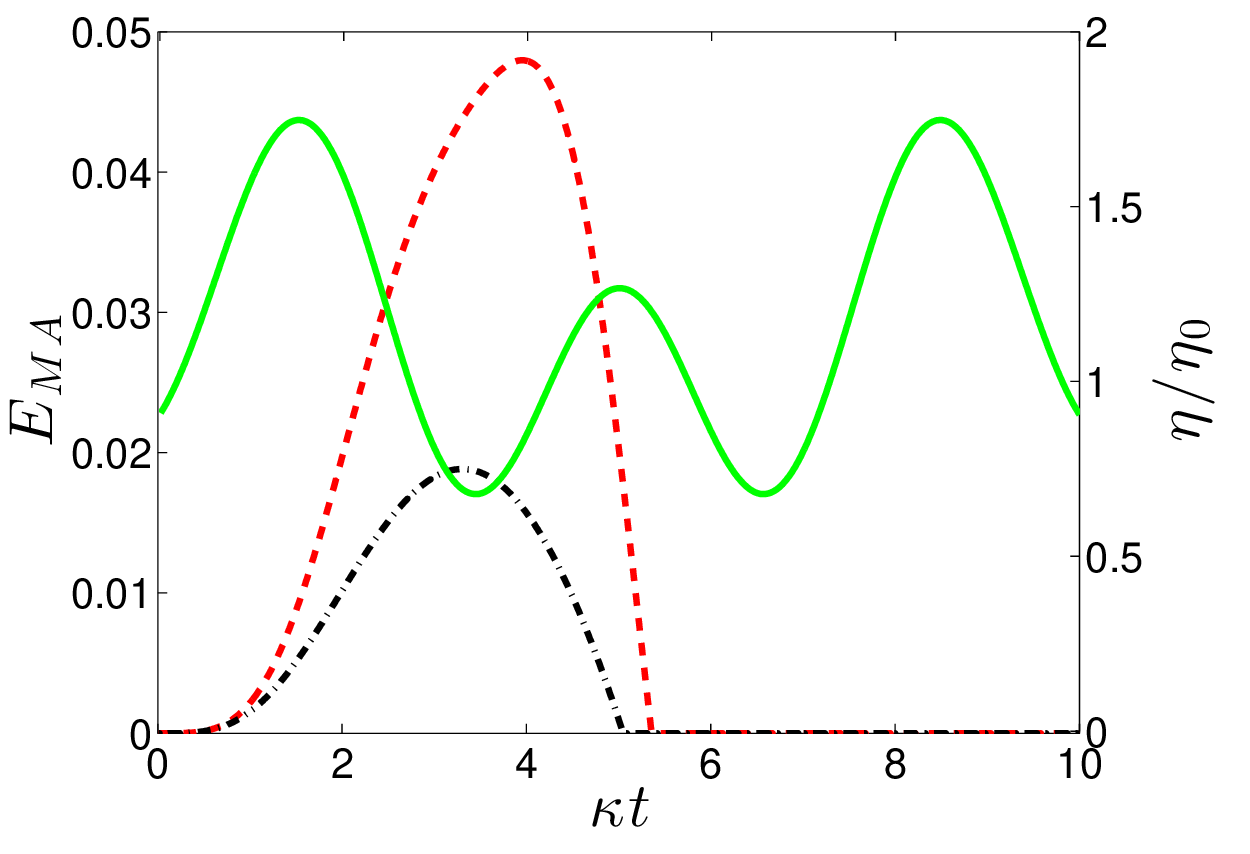}
\caption{(Color online) Entanglement dynamics $E_{MA}(t)$ (dashed line) with the optimal laser intensity modulation $\eta(t)$ (solid line). The entanglement $E_{MA}(t)$ for constant $\eta(t)=\eta_0$ is also shown (dashed-dotted line).  All plotted units are dimensionless.}
\label{fig:optimal}
\end{center}
\end{figure}

 Inspired by the techniques for dynamical optimization proposed in~\cite{doria}, we call $\eta_0$ the unmodulated value of $\eta$ and take
\begin{equation}
\eta(t) = \eta_0 + \sum_{j=1}^{j_{max}}\left[ A_j \cos(\omega_j t)+B_j\sin(\omega_j t) \right],
\end{equation}
where $\omega_j = 2\pi j/\tau+\delta_j$ are the harmonics and $\delta_j$ is a small random shift. The coefficients $A_j$ and $B_j$ are chosen in a way that the total energy brought about by the time-modulated field is the same as the one associated with the unmodulated case.
We then set the time-window so that $\tau=3.4\kappa^{-1}$, when we observe the maximum value of $E_{MA}$ in the unmodulated instance. The other parameters are as in Fig.~\ref{fig:deltaplots}. We then look for the parameters $A_j$ and $B_j$ that  maximise the value of $E_{MA}(\tau)$ for a given set of random shifts $\delta_j$. We use standard optimisation routines to find a (local) maximum of $E_{MA}(\tau)$. We repeat the search of the optimal coefficients  for different values of $\delta_j$ and take the overall maximum. The corresponding results are presented in Fig.~\ref{fig:optimal}, where we show the optimal modulation $\eta(t)$ and the optimal $E_{MA}(t)$. These findings are also compared to the case without modulation. The maximum value attained in the interval $[0;\tau]$ is $E_{MA}(t)\simeq0.05$ which is 2.5 times larger than the case without modulation, thus demonstrating the effectiveness of our approach. 

\noindent
\subsection{Periodic modulation: long time entanglement}
\begin{figure}[t]
{\bf (a)}\hskip4cm{\bf (b)}
\includegraphics[width=\linewidth]{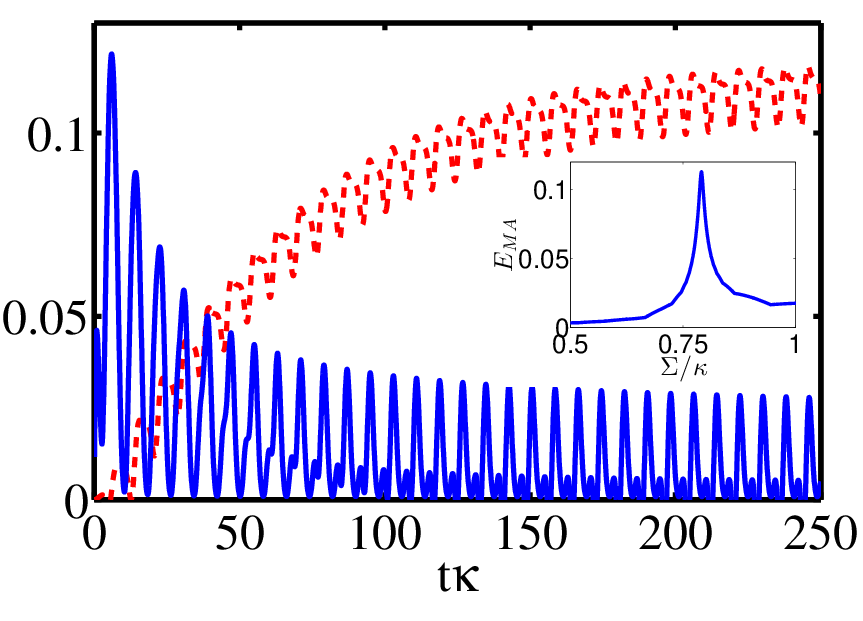}\\
\includegraphics[width=0.98\linewidth]{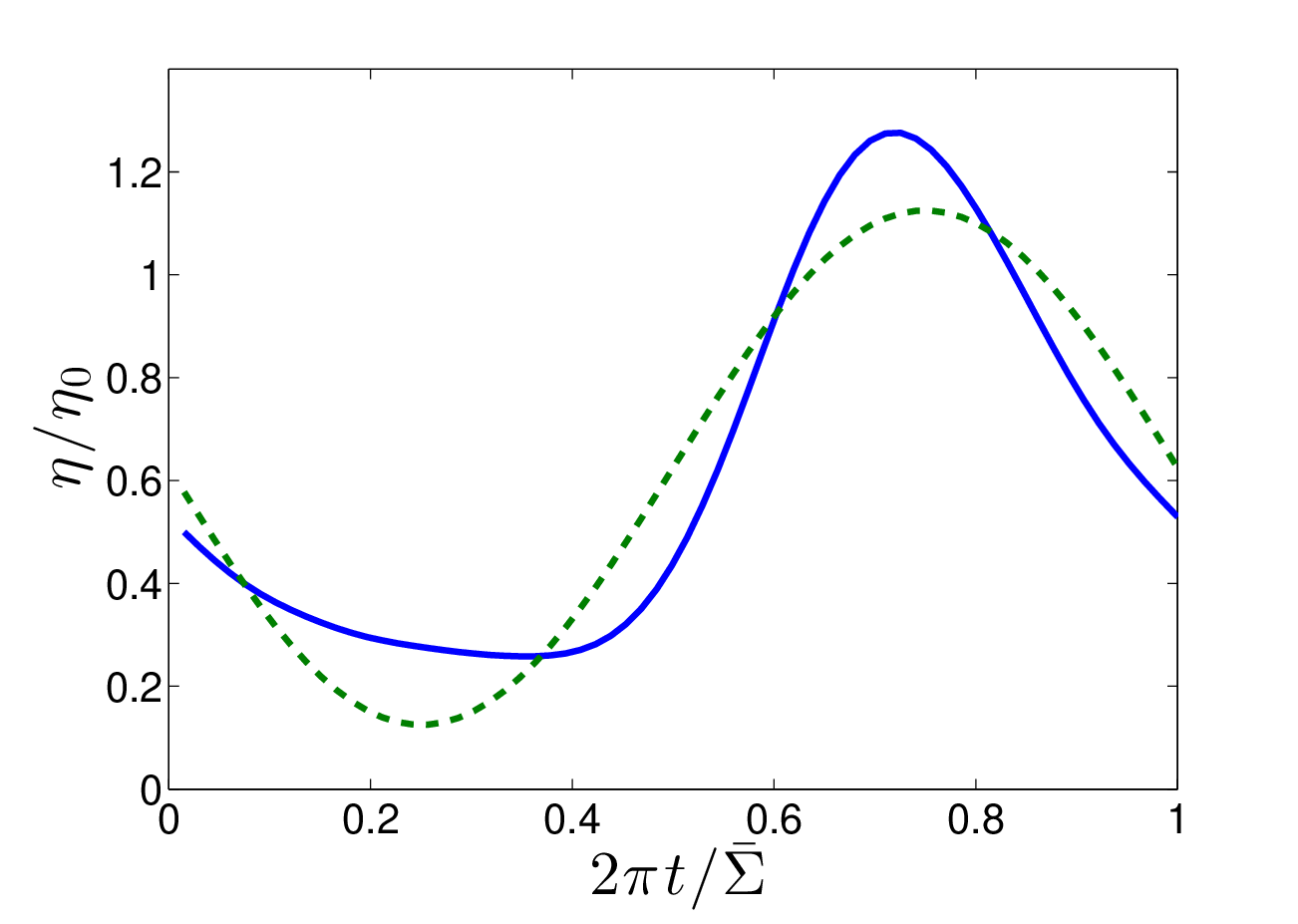}
\caption{\label{fig:periodic}
(Color online) {\bf (a)} Dynamics of the cavity-mirror and cavity-atoms entanglement $E_{CM,CA}$ (solid line) and atoms-mirror entanglement $E_{MA}$ (dashed line) for $\Sigma=\bar\Sigma\sim 0.79 \kappa$. Inset: Maximum $E_{MA}$ for long times with a periodic modulation as a function of the frequency $\Sigma$. {\bf (b)} Optimal periodic modulation $\eta(t)$ (solid line) for one period of time $2\pi/\bar\Sigma$ compared to the monochromatic modulation (dashed line).  All plotted units are dimensionless.}
\end{figure}

In the two situations analyzed so far [constant laser intensity $\eta$ and optimally modulated $\eta(t)$], the atom-mirror entanglement $E_{MA}(t)$ is destined to disappear at long times. A complementary approach based on a periodic modulation of the laser intensity $\eta(t)$ was used in the pure optomechanical setting \cite{mari} to increase the long-time light-mirror entanglement. Here we use a similar approach by assuming the monochromatic modulation of the laser-cavity coupling
%
$\eta(t) =\eta''_0+{\eta'_0}\left[1 - \sin(\Sigma t)\right]$,
%
where $\Sigma$ is the frequency of the harmonic modulation, $\eta'_0=4\eta''_0={\eta_0}/2$, $\eta_0$ being the same constant coupling parameter taken before. These choices ensure that the approximations used in the dynamical analysis are valid. After the transient dynamics, the covariance matrix and, in turn, $E_{MA}(t)$ become periodic functions of time. In order to achieve the best possible performance at long times, we compute the maximum of $E_{MA}(t)$ after the transient behavior, scanning the values of $\Sigma$. The result is shown in  Fig.~\ref{fig:periodic} {\bf (a)} (inset) revealing a sharp resonance with a maximum value of $E_{MA}\simeq 0.12$ for $\Sigma=\bar\Sigma\sim 0.79 \kappa$ (no further peak appears beyond this interval).  This arises as a result of the effective interaction between atoms and mirror mediated by the cavity field. As shown in Fig.~\ref{fig:deltaplots}{\bf (a)} the entanglement dynamics strongly depend on the effective detuning giving rise to such optimal behavior. Similar results have also been observed in Ref~\cite{mari}.
The analysis of the evolution of $E_{CM(CA)}$ and $E_{MA}$, shown in the main panel of Fig.~\ref{fig:periodic} {\bf (a)}, reveals that while $E_{CM(CA)}$ develops very quickly due to the direct cavity-atoms and cavity-mirror couplings, $E_{MA}$ grows in a longer time lapse, during which the cavity disentangles from the dynamics. The quasi-asymptotic value achieved by $E_{MA}$ reveals a sixfold increase with respect to the unmodulated case. This behavior is worth commenting as it strengthens our intuition that any atom-mirror entanglement has to result from a process that effectively couples such subsystems, bypassing any mechanism giving rise to multipartite entanglement within the overall system. 
Finally, we discuss the results achieved in the long time case by adopting an optimal-control technique similar to the one used for enhancing the short time entanglement. We have considered the periodic modulation of the intensity at the frequency $\bar\Sigma$ given by
\begin{equation}
\label{eq:mod}
\eta(t) = {\eta''_0}+{\eta'_0}\left[1 - \sum_{n=1}^{n_{max}} A_n\sin(n\bar\Sigma t)+B_n\cos(n\bar\Sigma t)\right],
\end{equation}
and looked for the coefficients $\{A_n,B_n\}$ optimizing $E_{MA}(t)$ at long times with the constraint: $\sum_{n=1}^{n_{max}}( A_n^2+B_n^2)\le 1$, ensuring that no instability is introduced in the dynamics of the overall system. In our simulation, $n_{max}=8$ has been taken to limit the complexity of the modulated signal. The resulting optimal coefficients give the periodic modulation shown in Fig.~\ref{fig:periodic} {\bf (b)} and a maximum entanglement $E_{MA}\simeq0.17$ which is about $30\%$ larger than the results obtained for the monochromatic modulation. This demonstrates the powerful nature of our scheme. Our extensive multiple-harmonic  analysis is able to outperform quite significantly the single-frequency driving scheme addressed above and discussed in Ref.~\cite{mari}, proving strikingly the sub-optimality of the monochromatic modulation, both at short and, surprisingly, at long times of the system dynamics. This legitimates  experimental efforts directed towards the use of time-modulated driving signals for the optimal control of the hybrid mesoscopic systems addressed here and similar ones based, for instance, on the use of a vibrating membrane~\cite{thompson} or a levitated nano-sphere~\cite{sphere} instead of the BEC.  

It should be noted that an adiabatic approach is used in Ref.~\cite{mari} to find an effective Hamiltonian.  When performed in our hybrid optomechnical scheme, such technique would remove the dynamics of the Bogoliubov modes of the atomic subsystem.

We consider the robustness of our protocol with respect to inaccuracy in the control of the value $\chi=\xi$ as follows: after finding the best periodic modulation assuming $\chi=\xi$, we ran again the simulations with the same modulation with $\chi=1.1 \xi$ and $\chi=0.9 \xi$. These values correspond to a 10\% inaccuracy in the nominal values of $\chi$ and $\xi$. We found that the maximum entanglement is at most only 3\% less than the original value thus confirming the stability of our result. Notice also that if the imbalance between $\chi$ and $\xi$ is known, for example by a calibration measurement, we can in principle run the optimization including the actual values of $\chi$ and $\xi$ therefore aiming at a larger entanglement value.

\noindent
\section{Conclusions} 
We have demonstrated that the modulation-assisted driving of a hybrid optomechanical device gives rise to interesting and rich entanglement dynamics, surpassing the limitations associated with a steady-state analysis and a constant pump. A significant improvement in the genuinely mesoscopic entanglement between the mode embodied by the atomic system and the mechanical one can be achieved by pumping the cavity with a modulated driving field, in both the short-time case and the long-time case. We have shown the existence of modulations that are able to beat the performance of simple monochromatic driving in terms of the maximum entanglement created between the mirror and the atomic system, thus favoring the creation of entanglement at long times. 

Our study strengthens the idea that important advantages are in order when optimal control techniques are implemented in the open-system dynamics of mesoscopic devices. This contributes to the current quest for the grounding of such approaches as valuable instruments for the control of mesoscopic (and multipartite) systems of various realizations, including interesting configurations of current experimental interest~\cite{thompson,sphere} to which our framework can be fully applied.

\acknowledgements

We thank M. Genoni, V. Giovannetti, and A. Xuereb for fruitful discussions. This work is supported by the UK EPSRC through a Career Acceleration Fellowship and a grant under the ``New Directions for EPSRC Research Leaders" initiative (EP/G004759/1).

\end{document}